\documentclass[aps,prl,preprintnumbers,epsf]{revtex4}
\newcommand{\gtap}{\;{\raise.3ex\hbox{$>$\kern-.75em\lower1ex\hbox{$\sim$}}}\;}
\newcommand{\ltap}{\;{\raise.3ex\hbox{$<$\kern-.75em\lower1ex\hbox{$\sim$}}}\;}
\usepackage[dvips]{graphicx}
\newcommand{\bea}{\begin{eqnarray}}
\newcommand{\eea}{\end{eqnarray}}

\begin{document}
\preprint{nucl-th/0409023v4}
\title{Renormalization of $NN$ scattering: contact potential}
\author{Ji-Feng Yang and Jian-Hua Huang}
\address{Department of Physics,
East China Normal University, Shanghai, 200062, China}
\date{\today}

\begin{abstract}
The renormalization of the $T$-matrix for $NN$ scattering with a
contact potential is reexamined in a nonperturbative regime
through rigorous nonperturbative solutions. Based on the
underlying theory, it is shown that the ultraviolet divergences in
the nonperturbative solutions of the $T$-matrix should be
subtracted through "endogenous" counter terms, which in turn leads
to a nontrivial prescription dependence. Moreover, employing the
effective range expansion, the importance of imposing physical
boundary conditions to remove the nontrivial prescription
dependence, especially before making physical claims, is discussed
and highlighted. As byproducts, some relations between the
effective range expansion parameters are derived. We also discuss
the power counting of the couplings for the nucleon-nucleon
interactions and other subtle points present within EFT framework
beyond perturbative treatment.
\end{abstract}
\pacs{03.65.Nk;11.10.Gh} \maketitle
\section{introduction}
The effective field theory (EFT) method or strategy~\cite{EFFT}
has become a primary tool for studying a variety of low energy
problems in particle and nuclear physics; important examples
include chiral perturbation theory ($\chi$PT)\cite{ChPT}, heavy
quark effective theory (HQET)\cite{HQET}, and non-relativistic
quantum chromodynamics (NRQCD)\cite{NRQCD}. As an EFT parametrizes
the short distance physics in a simple way, severe ultraviolet
(UV) divergences appear. Then one must carefully work out
pertinent power counting rules and renormalization
prescriptions\cite{NRQCD2}. In nonperturbative regimes (for
example, in the application of $\chi$PT to low energy nucleon
systems as is advocated by Weinberg\cite{WeinEFT}), the
'interplay' between power counting schemes and renormalization
prescriptions becomes quite complicated\cite{BevK}. To establish a
more reasonable and consistent framework, many proposals have been
put forward\cite{BevK,New,Nieves,VA}, creating controversies still
to be settled. There were once (perhaps are) even doubts about the
applicability of the EFT method.

The main difficulties stem from a distinct feature of the
nonperturbative formulation, which invalidates the naive use of
the perturbative renormalization (through subtraction)
programs\cite{JFY,YangPRD}. In this report, we continue our
investigations of the renormalization of the EFT for
nucleon-nucleon scattering in the nonperturbative regimes, which
we started in Ref.\cite{JFY}. We work here with a contact
potential that allows us to rigorously obtain a closed form of the
$T$-matrix through the use of the Lippmann-Schwinger equation
(LSE)\cite{Phillips}. In this way, many new features arising in
the nonperturbative regime can be explicitly illustrated. In our
approach, we utilize the underlying theory to understand the
renormalization of an EFT. In this work we will examine the
renormalization prescription dependence of the on-shell $T$-matrix
together with the observables or parameters coming from the low
energy theorems for nucleon-nucleon scattering. The latter are
obtained through the effective range expansion\cite{Cohen}. This
paper is organized as follows: In Sec. II, we sketch the
nonperturbative parametrization of the $T$-matrix proposed in
Ref.\cite{JFY} and its implications for nonperturbative
renormalization. In Sec. III, we employ the algebraic method
described in Ref.\cite{Phillips} to obtain a rigorous closed form
solution of the LSE in the case of contact nucleon-nucleon
interactions. Then the regularization and renormalization of the
$T$-matrix in the nonperturbative regime are analyzed using the
closed form solutions at various chiral orders. Both the
prescription dependence and its removal from the observables or
parameters obtained via the effective range expansion (low energy
theorems) are investigated in Sec. IV. In Sec. V, we study the
interplay between the renormalization prescription and the power
counting schemes for EFT, and the renormalization group evolution
in the nonperturbative regime and the naturalness of the
$T$-matrix are also explored. Finally, Sec. VI contains our
summary. Our main conclusion is that one should be aware of the
nontrivial renormalization prescription dependence in the
nonperturbative regime, with emphasis on physical boundaries.

\section{A Compact parametrization}
Let us start with a standard parametrization for the on-shell
partial wave $T$-matrix\cite{BevK} (we consider the diagonal
channels for simplicity):
\begin{eqnarray}
\label{para}
T_{l;os}(p)=-\frac{4\pi}{M}\frac{1}{p\cot\delta_l(p)-i p},
\end{eqnarray}with $M$ and $p$ being, respectively, the mass and
on-shell momentum of a nucleon, and $l$ denoting the angular
momentum number. Literally, the potential could be systematically
constructed or calculated using $\chi$PT\cite{WeinEFT} through
counting the powers in terms of $p^2/\Lambda^2$ or
$m_{\pi}^2/\Lambda^2$ with $\Lambda$ being the high scale or upper
limit for the EFT under consideration
($\Lambda\sim500\texttt{MeV}$). Then the off-shell $T$-matrix for
partial wave $l$ could be found through the solution of LSE,
\begin{eqnarray}\label{LSE} &&T_l(p^{\prime},p;
E)=V_l(p^{\prime},p)+\int \displaystyle\frac{kdk^2}{(2\pi)^2}
V_l(p^{\prime},k) G_0(k;E^+)
T_l(k,p; E),\\
&&G_0(k;E^+)\equiv \frac{1}{E^{+}-k^2/M},
\end{eqnarray} where $E^{+}\equiv E + i \epsilon$ with $E$ being
the center of mass energy. It should be noted, that
Eq.(~\ref{LSE}) is ill defined, as the potential $V_l$ calculated
using $\chi$PT is usually singular.

To highlight the nonperturbative features present in this problem,
we previously proposed a tentative nonperturbative parametrization
for the $T$-matrix based on LSE\cite{JFY},
\begin{eqnarray}
 \label{npt}
&&\frac{1}{T_l(p,p^{\prime};E)}=\frac{1}{V_l(p,p^{\prime})}-
{\mathcal{G}}_l(p,p^{\prime};E), \\
\label{G} && {\mathcal{G}}_l (p,p^{\prime};E)\equiv
 \frac{\int \frac{kdk^2}{(2\pi)^2} V_l
(p^{\prime},k ) G_0(k;E^+) T_l(k,p; E)}{
V_l(p,p^{\prime})T_l(p,p^{\prime};E)},
\end{eqnarray} where ${\mathcal{G}}_l$ carries the
nonperturbative information of all the quantum processes (all the
loop amplitudes in the field theoretic terminology) generated by
$V_l$. We have already shown in \cite{JFY} that the
nonperturbative quantity ${\mathcal{G}}_l$  could not be
renormalized through the introduction of "exogenous" counter terms
in the potential\cite{exo}. That means ${\mathcal{G}}$, or
equivalently $T$, is regularization and renormalization (R/R)
prescription dependent. There can be only one renormalization
prescription (contrary to perturbative treatment\cite{scheme})
consistent with physical data or boundary conditions. In the
following sections we will demonstrate this point with rigorous
solutions of LSE.
\section{Contact potential: rigorous solutions}
The specific case of contact potential allows us to transform the
integral equation into the algebraic one following
Ref.\cite{Phillips}. In other words, the nonlocal pion exchange
contributions to the potential are neglected. However, the main
conclusions remain qualitatively valid also when pion exchanges
are included.
\subsection{Factorized LSE and its algebraic solutions}
Next, to illustrate the role played by the nonperturbative
features, we will employ the $^1S_0$ channel. To see how the
situation evolves with the chiral orders, we consider the solution
of LSE with the following local potentials at three different
chiral orders ($\Delta=0 ,2 ,4 )$ (leading order, next-to-leading
order, next-to-next-to-leading order):
\begin{eqnarray}
\label{localV1} &&\Delta=0: V^{^1S_0}_{(0)}=C_0;\\\label{localV2}
&&\Delta=2: V^{^1S_0}_{(2)}= C_0 +C_2
(p^2+{p^{\prime}}^2);\\
\label{localV3}&&\Delta=4: V^{^1S_0}_{(4)}=C_0 +C_2
(p^2+{p^{\prime}}^2)+\tilde{C}_4p^2{p^{\prime}}^2+C_4
(p^4+{p^{\prime}}^4).
\end{eqnarray}Following Ref.\cite{Phillips} we "factorize" the potentials
into matrices: $V=U^{\text{T}}\lambda U^{\prime}$ with $U$, and
$U^{\prime}$ being column vectors and $\lambda$ an $n\times n$
matrix. At next-to-next-to-leading order ($\Delta=4$), they
are\begin{eqnarray}
\label{lambda3}\lambda\equiv \left(%
\begin{array}{ccc}
  C_0& C_2 &C_4 \\
  C_2 & \tilde{C}_4 &0 \\
  C_4 &0 & 0 \\
\end{array}%
\right),\ \ \ U\equiv \left(%
\begin{array}{ccc}
  1, & p^2, & p^4 \\
\end{array}%
\right),\ \ \  U^{\prime}\equiv \left(%
\begin{array}{ccc}
  1, & {p^{\prime}}^2,& {p^{\prime}}^4 \\
\end{array}%
\right);
\end{eqnarray} whereas at next-to-leading order ($\Delta=2$) they read,
\begin{eqnarray}
\label{lambda2}\lambda\equiv \left(%
\begin{array}{cc}
  C_0& C_2 \\
  C_2 & 0 \\
  \end{array}%
\right),\ \ \ U\equiv \left(%
\begin{array}{cc}
  1, & p^2  \\
\end{array}%
\right),\ \ \  U^{\prime}\equiv \left(%
\begin{array}{cc}
  1, & {p^{\prime}}^2\\
\end{array}%
\right).
\end{eqnarray}The off-shell $T$-matrix factorizes exactly in the
same manner: $T=U^{\text{T}}\tau U^{\prime}$, where $\tau$ is an
$n\times n$ matrix. Generally, the coupling constants $[C_n]$ come
from the chiral expansion of an underlying theory (say, QCD) for
nucleons and pions in terms of $p^2/\Lambda^2$, and hence they
scale like $C_{2n}/C_0\sim \Lambda^{-2n} $ in the naive power
counting scheme.

Using this notation, the LSE can be reduced to the following
algebraic equation\cite{Phillips},
\begin{eqnarray}
\label{algebraicLSE}\tau(E^+)=\lambda+\lambda
\circ{\mathcal{I}}(E^+)\circ\tau(E^+)
\end{eqnarray}with $\circ$ denoting the matrix multiplication.
For the $3\times 3$ case, the matrix ${\mathcal{I}}$ and the
related parametrizations and definitions are listed in Appendix A.
With this algebraic parametrization, all the ill defined integrals
can be isolated and parametrized in any regularization
prescription. The solution to this algebraic equation is easy to
obtain:
\begin{eqnarray}
\tau(E^+)=(1-\lambda\circ {\mathcal{I}}(E^+))^{-1}\circ \lambda.
\end{eqnarray} In a similar fashion, the solution of the
$T$-matrix can be obtained. For the three chiral orders considered
so far, they read (on-shell),
\begin{eqnarray}
\label{Tn1} && \Delta=0:\frac{1}{T_{os}(p)}=\frac{1}{C_0}+ J_0
+\frac{M }{4\pi}ip; \\
\label{Tn2}&&\Delta=2:
\frac{1}{T_{os}(p)}=\frac{(1-C_2J_3)^2}{C_0+C_2^2
J_5+C_2(2-C_2J_3)p^2}+ J_0 +\frac{M }{4\pi}ip;\\
\label{Tn3}&& \Delta=4:\frac{1}{T_{os}(p)}=\frac {N_0+N_1 p^2+N_2
p^4}{D_0+D_1p^2+D_2p^4+ D_3 p^6}+ J_0 +\frac{M }{4\pi}ip,
\end{eqnarray}
with the ill defined integrals [$J_n$] and the coefficients
$[N_n]$ and $[D_n]$ being defined in Appendix B. The
next-to-leading order has been considered in Ref.\cite{Phillips},
whereas the $\Delta=4$ result has not been given before. We should
note that $J_0$ always stands "alone" in the real part of the
inverse on-shell $T$-matrix. (A rigorous proof of this fact at any
order is given in Appendix C.) Here we would like to stress the
compact or closed form of the expressions for the $T$-matrix in
terms of the couplings and the integrals [$J_n$]. It is this
crucial property that distinguishes the nonperturbative solutions
from the perturbative ones and complicates the renormalization.

Obviously, the $T$-matrix becomes more complicated as higher order
interactions are included. Nevertheless, when the nonlocal pion
exchanges are included, one can naturally anticipate that the
nonperturbative solutions still take {\em compact or closed}
forms. Thus we expect, that, our conclusions here will also hold
in the realistic potentials with nonlocal pion contributions, at
least qualitatively.

\subsection{Failure of 'exogenous' counter-term renormalization}
Now let us consider the renormalization of the $T$-matrix. The
leading order case is trivial; one can absorb the only divergence
in $J_0$ into the inverse coupling: $1/C_0$, similar to the
perturbative cases.

However, in the presence of higher order interactions, such
operation may not work. For example, at next-to-leading order, in
order to renormalize the on-shell $T$-matrix in Eq.(\ref{Tn2}), or
to make the fraction $ \frac{(1-C_2J_3)^2}{C_0+C_2^2
J_5+C_2(2-C_2J_3)p^2}+ J_0  $ finite, one should make each of the
following compact functions finite at the same time:
\begin{eqnarray}\label{compactfrac}
(C_0+C_2^2J_5)/(1-C_2J_3)^2,\ \ C_2(2-C_2J_3)/(1-C_2J_3)^2, \ \
J_0.
\end{eqnarray} Now it is clear that the main obstacle for
performing "exogenous" subtraction for the $T$-matrix is the
compact or closed expressions in terms of $[C_n]$ and
$[\bar{J}_n]$: it is hard to see how to make
$(C_0+C_2^2J_5)/(1-C_2J_3)^2,\ \ C_2(2-C_2J_3)/(1-C_2J_3)^2$ and
$J_0$ finite simultaneously, since each of them is a compact or
closed expression expressed given in terms of the two couplings,
$C_0,C_2$, and the three divergent integrals, $J_0,J_3$ and $J_5$.
The situation differs strikingly from the perturbative case where
counter terms are introduced order by order with the higher order
terms discarded, as no compact or 'closed' form of expression is
involved. Moreover, no matter what was done for
$(C_0+C_2^2J_5)/(1-C_2J_3)^2$ and $C_2(2-C_2J_3)/(1-C_2J_3)^2$,
one should make sure that $J_0$ stays "separately" finite at the
same time.

\subsection{Nonperturbative renormalization in EFT}
It is known that an EFT is often established through certain
reorganization of parts of a well defined underlying theory (UT,
at least renormalizable). Unfortunately, such reorganization
usually (1) brings about new UV divergences and (2) impedes the
exogenous counter terms from working. To see the first point,
consider the diagrams shown in Figs.1 and 2, with the heavy meson
exchange diagrams (with $g$ and $m_h$ being the coupling constant
and the meson mass) underlying the ones with contact interactions.
For convenience, let us introduce a projection operator
$\breve{{\mathcal{P}}}_{\text{\tiny LE}}$ to symbolize the
influences of this heavy meson: the extraction of the EFT vertices
or couplings from the UT diagrams.

At tree level (Fig.1),\bea
-i\bar{C}_0\equiv\breve{{\mathcal{P}}}_{\text{\tiny
LE}}\Gamma^{(4)}_{tree}=\breve{{\mathcal{P}}}_{\text{\tiny LE}}
\{\frac{-ig^2}{k^2-m^2_h}\}=i\frac{g^2}{m_h^2},\eea no divergence
appears. The complication comes at the loop diagram level. For
example, in the case of the convergent box diagram in Fig.2, if
$\breve{{\mathcal{P}}}_{\text{\tiny LE}}$ is applied after the
loop integration ($\int \frac{d^4l}{(2\pi)^4}$) has been done
({\em the correct order}), one would get a well defined expansion
in terms of $\frac{1}{m_h^2}$. When
$\breve{{\mathcal{P}}}_{\text{\tiny LE}}$ is applied before $\int
\frac{d^4l}{(2\pi)^4}$ ({\em the incorrect order}), the divergent
bubble diagram results. Thus the new divergences generally arise
from the incorrect order of computations, as the following
commutator does not vanish identically: \bea \label{CT}
\hat{O}_{\texttt{c.t.}}\equiv[\breve{{\mathcal{P}}}_{\text{\tiny
LE}},\int \frac{d^4l}{(2\pi)^4}] \neq0.\eea Embarrassingly, one
has to use EFT either because UT is unavailable or because the
calculations in UT are tedious. Combining this procedure with
nonperturbative context (infinite iteration or resummation) makes
things even worse: the counter terms could not be implemented
exogenously.

However, from the underlying theory the solution follows
immediately: one should devise some procedures to effectively
"recover" the correct order for
$\breve{{\mathcal{P}}}_{\text{\tiny LE}}$ and $\int
\frac{d^4l}{(2\pi)^4}$ before anything else is done. The clue lies
in Eq.(\ref{CT}). Through rearrangement, Eq.(\ref{CT}) is
equivalent to the following equation for the integrand of a loop
diagram (say, the box diagram integrand $f_{box}$), \bea
\label{CTsubtrac} \breve{{\mathcal{P}}}_{\text{\tiny LE}} \int
\frac{d^4l}{(2\pi)^4} f_{box}= \int
\frac{d^4l}{(2\pi)^4}\breve{{\mathcal{P}}}_{\text{\tiny LE}}
f_{box}+ \hat{O}_{\texttt{c.t.}} f_{box}=\int
\frac{d^4l}{(2\pi)^4} f_{bubble}+ \hat{O}_{\texttt{c.t.}}
f_{box}.\eea That means, in order to recover the correct-order
results in EFT, we must introduce a counter term:
$\hat{O}_{\texttt{c.t.}} f_{box}$. Therefore, the UT scenario
provides a natural interpretation for the counter terms, and more
importantly, a rationality for the subtraction at the level of the
loop integral {\em without} any reference to Lagrangian. That is,
the counter terms must be endogenous: the divergent integrals must
be subtracted {\em before} the nonperturbative
reorganization\cite{JFY}. Finally, the subtracted integrals
(finite) appear in the compact nonperturbative expressions which
are no longer compatible with exogenous counter terms. In this
logic, the formal consistency issue\cite{BevK} of the Weinberg
power counting simply dissolves: in the nonperturbative regime of
EFT, there is no point in searching for exogenous counter terms
and their counting rules. Of course there might be other
approaches that directly renormalize the integrals without the
explicit use of counter terms.

Thus the nonperturbative renormalization must be implemented
either through endogenous counter terms or other means that
effectively "subtract" the EFT integrals (see similar
prescriptions in Ref.\cite{Ge_Niev}). For the $T$-matrix
considered in this paper, this procedure would be formulated as a
simple replacement of the divergent $[J_n]$ with the subtracted
$[\bar{J}_n]$, which are finite constants (prescription dependent
and hence arbitrary), in the compact expressions:
\begin{eqnarray}
\label{Tn1r} && \Delta=0: \bar{T}_{os}^{-1}(p)=\frac{1}{C_0}+
\bar{J}_0
+\frac{M }{4\pi}ip;\\
\label{Tn2r}&&
\Delta=2:\bar{T}_{os}^{-1}(p)=\frac{(1-C_2\bar{J}_3)^2}{C_0+C_2^2
\bar{J}_5+C_2(2-C_2\bar{J}_3)p^2}+ \bar{J}_0 +\frac{M }{4\pi}ip;\\
\label{Tn3r}&& \Delta=4: \bar{T}_{os}^{-1}(p)=\frac
{\bar{N}_0+\bar{N}_1 p^2+\bar{N}_2
p^4}{\bar{D}_0+\bar{D}_1p^2+\bar{D}_2p^4+ \bar{D}_3 p^6}+
\bar{J}_0 +\frac{M }{4\pi}ip.
\end{eqnarray} Here we wish to note that from a UT perspective,
both $[C_n]$ and $[\bar{J}_n]$ come from the projection acting on
the convergent vertices in UT. (Those that are divergent in UT
will be renormalized before applying the projection and do not
directly contribute to EFT renormalization due to scale hierarchy,
we will return later to this point in Sec. VI.) In this sense,
$[\bar{J}_n]$ are also fundamental parameters in EFT, so the
nonperturbative $T$-matrix is parametrized by both $[C_n]$ and
$[\bar{J}_n]$. To illustrate this point, let us apply the
projection on the box diagram after the loop integration is
carried out. After some calculations we get
\bea\breve{{\mathcal{P}}}_{\text{\tiny
LE}}\Gamma^{(4)}_{box}|_{\texttt{leadingterm}
}=-i\frac{g^4}{m_h^4}I^{\text{\tiny (UT)}}_0(M,m_h,p)
=i{\bar{C}}^2_0[J^{\text{\tiny (UT)}}_0(M,m_h)+\frac{M
}{4\pi}ip],\eea where the definite parameter $J^{\text{\tiny
(UT)}}_0(M,m_h)$ (see Appendix B), in a place of the divergent
integral $J_0$, can be extracted in the following way\bea
J^{\text{\tiny (UT)}}_0(M,m_h)=-\frac{{\texttt{Re}}(i
\breve{{\mathcal{P}}}_{\text{\tiny
LE}}\Gamma^{(4)}_{box})}{(i\breve{{\mathcal{P}}}_{\text{\tiny
LE}}\Gamma^{(4)}_{tree})^2}|_{p=0}=-\breve{{\mathcal{P}}}_{\text{\tiny
LE}}\left \{\frac{{\texttt{Re}}(i
\Gamma^{(4)}_{box})}{(i\Gamma^{(4)}_{tree})^2}\right\}|_{p=0}.\eea

We should note that here we used the mesonic interaction for
illustration. Of course, the true contact nucleon interactions
should be computed from QCD. But the mechanism explained above
still holds in general.

Finally, we note that the renormalized $T$-matrix suffers from
severe prescription dependence in the nonperturbative regime,
which is incompatible with the exogenous counter terms. That
means, given specific couplings, only one prescription could yield
the physical $T$-matrix; others have to be dropped even though
they are finite. So the final resolution boils down to the
flexible regularization methods that could facilitate convenient
access to physical predictions\cite{JFY}, as was already noted in
other nonperturbative contexts\cite{nptreg}. This argument leads
us to the following strategy: One first parametrizes the ill
defined integrals in terms of ambiguous constants and then imposes
physical boundary conditions. Similar strategy also based on the
underlying theory, has already been described in Ref.\cite{YYY}
for renormalizing any EFT.

\section{low energy theorems (LET) and
prescription dependence} \subsection{Effective range expansion
}Now let us consider effective range expansion (ERE) defined as
follows,
\begin{eqnarray}
\label{ERT}\texttt{Re}\left \{-\frac{4\pi}{M}T^{-1}_{os}(p)\right
\}=p\cot\delta(p)=-\frac{1}{a}+\frac{1}{2}r_e
p^2+\sum_{k=2}^{\infty} v_k p^{2k},
\end{eqnarray}with the parameters $a$ and $r_e$ being the
scattering length and the effective range, which (including
$[v_k]$) could be extracted from the scattering data. In this
sense, we could impose their values as the boundary conditions for
the $T$-matrix. Performing the expansion for the $T$-matrix
obtained above, we get:\begin{eqnarray}
\Delta=0:p\cot\delta(p)&&=-\frac{4\pi}{M}\left \{C_0^{-1}+
\bar{J}_0 \right \};\\
\Delta=2:p\cot\delta(p)&&=-\frac{4\pi}{M}
\left\{\bar{\nu}_0{\bar{\delta}_0}^{-1}+\bar{J}_0-
\bar{\nu}_0\bar{\delta}_1{\bar{\delta}_0}^{-2}p^2 +
\sum_{k=2}^{\infty}
\bar{\nu}_0\bar{\delta}_1^k{\bar{\delta}_0}^{-k-1}(-p^2)^k\right
\}, \nonumber \\&&\bar{\nu}_0\equiv (1-C_2\bar{J}_3)^2,\ \
\bar{\delta}_0\equiv C_0+C_2^2
\bar{J}_5,\ \ \bar{\delta}_1\equiv C_2(2-C_2\bar{J}_3);  \\
\Delta=4:p\cot\delta(p)&&=-\frac{4\pi}{M}\left
\{\bar{N}_0{\bar{D}_0}^{-1}+(\bar{N}_1\bar{D}_0-
\bar{N}_0\bar{D}_1){\bar{D}_0}^{-2}p^2 \right.
\nonumber\\
&& +[\bar{N}_2\bar{D}_0^2- \bar{N}_1\bar{D}_1\bar{D}_0
+\bar{N}_0(\bar{D}_1^2-\bar{D}_0\bar{D}_2)]\bar{D}_0^{-3}p^4\nonumber
\\ &&+[\bar{N}_0(2\bar{D}_1\bar{D}_2\bar{D}_0-
\bar{D}_3\bar{D}_0^2 -\bar{D}_1^3)
+\bar{N}_1\bar{D}_0(\bar{D}_1^2-\bar{D}_0\bar{D}_2)
-\bar{N}_2\bar{D}_1\bar{D}_0^2]\bar{D}_0^{-4}p^6\nonumber \\
&&\left. + \cdots\right \}.
\end{eqnarray}The scattering length, effective range, and the $v_k$
can be read from the results above. For the three orders
considered so far, we have\begin{eqnarray}\label{ERE0} \Delta=0:
&& a^{-1}=\frac{4\pi}{M}(C_0^{-1}+\bar{J}_0),\ \ r_e=0,\ \ v_k=0,
k\geq2;\\\label{ERE2} \Delta=2: &&
a^{-1}=\frac{4\pi}{M}(\bar{\nu}_0{\bar{\delta}_0}^{-1}+\bar
{J}_0),\ \ r_e= \frac{8\pi}{M}\bar{\nu}_0\bar{\delta}_1
{\bar{\delta}_0}^{-2},\ \ v_k=
\frac{4\pi}{M}\bar{\nu}_0\bar{\delta}_1^k{(-\bar{\delta}_0)}^{-k-1},
k\geq 2; \\\label{ERE4} \Delta=4: &&
a^{-1}=\frac{4\pi}{M}(\bar{N}_0{\bar{D}_0}^{-1}+\bar {J}_0),\ \
r_e=\frac{8\pi}{M}
(\bar{N}_0\bar{D}_1-\bar{N}_1\bar{D}_0)\bar{D}_0^{-2},\nonumber
\\&& v_2=\frac{4\pi}{M}[\bar{N}_0(\bar{D}_0\bar{D}_2-\bar{D}_1^2)+
\bar{N}_1\bar{D}_1\bar{D}_0
-\bar{N}_2\bar{D}_0^2]\bar{D}_0^{-3},\nonumber\\
&&v_3=\frac{4\pi}{M}[\bar{N}_0(\bar{D}_3\bar{D}_0^2-
2\bar{D}_1\bar{D}_2\bar{D}_0 +\bar{D}_1^3)
-\bar{N}_1\bar{D}_0(\bar{D}_1^2-\bar{D}_0\bar{D}_2)
+\bar{N}_2\bar{D}_1\bar{D}_0^2]\bar{D}_0^{-4},\\
 &&\cdots \nonumber
\end{eqnarray}Note again that, at each order, $\bar{J}_0$ only
enters the expression for the scattering length but is "decoupled"
with all the other ERE parameters. The reason is clear:
$\bar{J}_0$ stands alone in $T^{-1}$. As we make clear below, this
point has very important implications.

Keeping in mind that the parameters $[\bar{J}_n]$ are in principle
independent of each other, two distinct approaches can be adopted
in order to impose physical or reasonable boundary conditions: (1)
taking the couplings and the prescription parameters as the
fundamental variables, and the ERE parameters as the functions of
these variables; or (2) conversely, taking some of the ERE
parameters (which should be physical) as fundamental and the
others as the functions of them. For convenience, we could also
parametrize $[\bar{J}_n]$ in terms of a dimensional scale
${\widetilde{\mu}}$ and dimensionless numbers [$q_{\cdots}$]:
\begin{eqnarray}
\label{J_para} J_0 \equiv q_0 M \widetilde{\mu}, \ \ J_3\equiv
q_3M{\widetilde{\mu}}^3,\ \ J_5\equiv q_5M{\widetilde{\mu}}^5,\ \
J_7\equiv q_7M{\widetilde{\mu}}^7,\ \ J_9\equiv
q_9M{\widetilde{\mu}}^9.
\end{eqnarray}The appearance of $M$ is
easy to see from Appendix A. Generally, the magnitude of
${\widetilde{\mu}}$ could vary from a value on the EFT expansion
scale, $\Lambda$, to the value of the pion decay constant (much
smaller than $M$): $\widetilde{\mu}\in (f_{\pi},\Lambda)$. One can
also alter the integrals by letting the dimensionless numbers
$[q_n]$ to vary. Thus, a nonperturbative renormalization
prescription is parametrized by $[q_n;{\widetilde{\mu}}]$.
However, the magnitude of $[\bar{J}_n]$, which also comes from the
low energy projection in UT, should not be larger than the naive
powers of the chiral symmetry breaking scale $\Lambda_{\chi
SB}\simeq M$. In other words, we can safely assume that:
$|\bar{J}_n|\leq M^{n+1}, n\neq 0, |\bar{J}_0|\leq M^2.$

Having made these preparations, we can start to examine the low
energy expansions listed above order by order in chiral expansion.
The leading order case ($\Delta=0$) is trivial: we have only one
condition, i.e., imposing that $\frac{1}{C_0}+J_0=a$, with $a$
being experimentally measured, is enough since $r_e=v_k=0,
k\geq2$, which is obviously bad theoretical prediction, although
it is not prescription dependent. Thus, the situation at this
order is physically uninteresting, and most importantly, the
distinctive nonperturbative features we wish to expose are not
obvious here. Therefore we examine in detail the higher order
cases.
\subsection{LET at next-to-leading order: $\Delta=2$}
Let us start with the next-to-leading order: $\Delta=2$. As was
mentioned before, we shall discuss the problem from two
perspectives, respectively.
\subsubsection{First point of view}
For convenience, let us list the explicit expressions of $a, r_e$,
etc. in terms of $[C_n]$ and $[\bar{J}_n]$, which
read,\bea\label{sctlength1}
&&a=\frac{M}{4\pi}\frac{C_0+C_2^2\bar{J}_5}{(1-C_2
\bar{J}_3)^2+\bar{J}_0(C_0+C_2^2\bar{J}_5)},\nonumber \\
&&r_e=\frac{8\pi}{M}\frac{(2C_2-C_2^2\bar{J}_3)(1-
C_2\bar{J}_3)^2}{(C_0+C_2^2\bar{J}_5)^2},\nonumber \\
&&v_k=\frac{4\pi}{M}(-)^{k+1}\frac{(1-C_2J_3)^2C_2^k(2-C_2\bar{J}_3)^k}
{{(C_0+C_2^2 \bar{J}_5)}^{k+1}}, k\geq 2.\eea

Before imposing reasonable boundary conditions, we could not make
any physical predictions as $\bar{J}_0, \bar{J}_3$ and $\bar{J}_5$
each independently varies. The independent variations of
$\bar{J}_0, \bar{J}_3$ and $\bar{J}_5$ could not be easily
absorbed into the couplings due to the reasons explained in Sec.
III. Thus, unlike the leading order case, we need (more) exogenous
constraints to fix the values of $\bar{J}_0, \bar{J}_3$ and
$\bar{J}_5$.

From Eq.(\ref{sctlength1}), $\bar{J}_3$ and $\bar{J}_5$ could be
solved in terms of $r_e,v_2$ and $C_0,C_2$, the solution might be
unique after accounting for a reasonable size. After the insertion
of the obtained numbers back into the formula for $a$, $\bar{J}_0$
could be expressed in terms of the physical value of the
scattering length. In this sense, imposing two additional boundary
conditions could fix the prescription or make the next-to-leading
order result unambiguous. Now the [$v_k$] with $k\geq3$ are taken
to be theoretical predictions, which now shall be better than the
leading order ones, as we have more degrees of freedom to work
with: $\bar{J}_3$ and $\bar{J}_5$, which come together with the
new interactions. The fact, that some of the predictions are still
poor can be attributed to the inadequacy of the next-to-leading
order potential: even higher order terms should be put in and
accordingly more physical boundary conditions are needed. Thus, in
spite of the fact, that the procedure for fixing prescription
becomes more nontrivial and laborious, the predictions for the ERE
parameters improve when the higher order interactions are included
(because of the "freedom" brought by the augmented interactions).
We must repeat here that the predictions are made using the
prescription that is most compatible with the physical boundary
conditions.

Employing (\ref{J_para}), we could also write the above equations
as \bea \label{sctlength2} &&a=\frac{M}{4\pi}\frac{C_0+C_2^2q_5M
{\widetilde{\mu}}^5}{(1-C_2 q_3M{\widetilde{\mu}}^3)^2+
q_0M{\widetilde{\mu}}(C_0+
C_2^2q_5M{\widetilde{\mu}}^5)}, \nonumber \\
&&r_e=\frac{8\pi}{M}\frac{(2C_2-C_2^2q_3M{\widetilde{\mu}}^3)(1-
C_2q_3M{\widetilde{\mu}}^3)^2}{(C_0+ C_2^2q_5M{\widetilde{\mu}}^5)^2}, \nonumber\\
&&v_k=\frac{4\pi}{M}(-)^{k+1}\frac{(1-C_2q_3M{\widetilde{\mu}}^3)^2C_2^k(2-
C_2q_3M{\widetilde{\mu}}^3)^k} {{(C_0+C_2^2
q_5M{\widetilde{\mu}}^5)}^{k+1}}, k\geq 2.\eea Note that even
though the $\widetilde{\mu}$ scale dependence can be removed
(fixed), the prescription dependence remains in terms of the
dimensionless parameters $[q_n]$ that are independent of each
other, a subtle point that often seems to be overlooked. For
example, in a cutoff scheme, the renormalization is usually
performed in such a way that the cutoff dependence is removed by
letting the couplings to develop a certain cutoff dependence.
However, remains residual prescription dependence remains through
the cutoff independent but prescription dependent numbers $[q_n]$.
Without fully appreciating this point, any fitting procedure that
uses tuning of the couplings or even tuning of the cutoff only
amounts to fitting along a special orbit in the space $[q_n]$, not
in the whole space. The result thus obtained is still prescription
dependent.

\subsubsection{Second perspective}
Taking $a$, $r_e$ and $\{v_2\}$ as the elementary parameters, we
can express all the higher order constants $[v_k,k\geq3]$ as
\begin{eqnarray}
\label{v_k} v_k=\frac{M-4\pi a\bar{J}_0}{Ma}(-2v_2/r_e)^k, k\geq3.
\end{eqnarray} Note that from this perspective the
prescription-dependent parameter $\bar{J}_0$ seems to be an
independent constant in addition to the three elementary
parameters. At first sight, this ambiguity calls for one more
condition: the value for $v_3$. But from the discussion above, we
know that this seemingly independent parameter is in fact
determined together with $\bar{J}_3$ and $\bar{J}_5$ by the
equations (\ref{sctlength1}). Of course the nature of the problem
remains the same even when it is taken as independent. The most
striking point here is again that the prescription is "removed"
through fixing, i.e., through boundary conditions, which is a
nontrivial procedure as articulated above.

One could also put Eq.(\ref{v_k}) into the form that contains no
explicit prescription dependence: \bea \label{v_kp}
v_k=v_3(-2v_2/r_e)^{k-3}, k\geq4. \eea This relation should hold
for any problems (in certain atomic or molecular contexts) with
contact potential such as $V(\textbf{x})\sim
C_{(0)}\delta^{(3)}(\textbf{x})+C_{(2)}\nabla^2\delta^{(3)}(\textbf{x})
$.

So, we may conclude that, no matter what point of view one adopts,
the key point is that the compact nonperturbative formulations
make the prescription dependence and its removal a very nontrivial
problem or procedure. However, the predictions also improve with
the use of these formulations, in spite of their technical
complexity. This nontrivial procedure will get more involved as
more higher order corrections to the potential are included. To
verify this, let us turn to the next-to-next-to-leading order.

\subsection{LET at next-to-next-to-leading order: $\Delta=4$}
Again we begin with the first perspective.
\subsubsection{First perspective}
Now we have four couplings ($C_0,C_2,C_4,\tilde{C}_4$) and five
prescription dependent parameters
($\bar{J}_0,\bar{J}_3,\bar{J}_5,\bar{J}_7,\bar{J}_9$) in eight
compact expressions:
$\bar{N}_0,\bar{N}_1,\bar{N}_2,\bar{D}_0,\bar{D}_1,\bar{D}_2,\bar{D}_3$
and $\bar{J}_0$, which stays alone. From Eq.(\ref{ERE4}) it is
clear that we need at least five conditions to fix
$\bar{J}_0,\bar{J}_3,\bar{J}_5,\bar{J}_7$ and $\bar{J}_9$, say
$a,r_e,v_2,v_3$ and $v_4$. But the compact expressions such as
that for the scattering length,
$a=\frac{M}{4\pi}\frac{\bar{D}_0}{\bar{N}_0 +\bar{J}_0 \bar{D}_0}$
with $ \bar{D}_0$ and $\bar{N}_0$ given in Appendix B, become more
involved. This means that the boundary conditions might be more
stringent for $\bar{J}_0,\bar{J}_3,\bar{J}_5,\bar{J}_7$ and
$\bar{J}_9$ and the analytical work more difficult. In the
meantime, the predictions for $v_k$ at this order should be better
than the leading and next-to-leading orders, as we have more
parameters.

Here some remarks are in order. At next-to-leading order, we
ignored the possible multiple solutions for the fixing procedure.
Here, with more compact expressions being involved, we should be
more careful about this multiplicity of solutions. To this end, we
note that the multiplicity could be effectively reduced with the
limitations on the reasonable magnitudes of $[\bar{J}_n]$ together
with the experimental values of the higher ERE parameters (say,
$v_k, k\geq 3$). However, no matter how the multiplicity is
removed, the solution is still an approximate (though
nonperturbative) one: The equations
(\ref{ERE0},\ref{ERE2},\ref{ERE4}) are obtained from a truncated
potential and could not be exact ones. Then, the theoretical
predictions based on such equations will be less credible,
especially for the ERE parameters that dominate higher and higher
energy regions. In other words, the boundary conditions should be
given by a procedure similar to fitting the shape of the phase
shift within the corresponding ranges at each chiral order. This
is actually what most authors have done, though the regularization
schemes used vary significantly. Of course our remarks in Sec.
IV.B.1 concerning the residual prescription dependence still apply
for all the higher order calculations.

Mathematically, the multiplicity of solutions might be generic for
nonperturbative renormalization because of the compact expressions
involved. Therefore the limit cycles encountered in the
Schr\"odinger approach of renormalizing singular
potentials\cite{limitcycle} might just be examples of such
multiplicity in certain regularization schemes.

\subsubsection{Second perspective}
To discuss the problem from the second perspective, we need to
express everything in terms of the first five ERE parameters. Then
the arguments go as in the preceding subsection. We shall not,
however, repeat such complicated technical details here. One could
also find the relations like Eq.(35) or (36), which hold true
independently of the prescriptions, by repeatedly using the
following recursive relations: \bea &&\tilde{v}_n=
-\sum_{k=1}^{3}\frac{\bar{D}_k}{\bar{D}_0} \tilde{v}_{n-k},
n\geq5;\ \ \tilde{v}_3= \frac{\bar{D}_3}{\bar{D}_0}\tilde{a}^{-1}-
\frac{\bar{D}_2}{\bar{D}_0}\tilde{r}_e-
\frac{\bar{D}_1}{\bar{D}_0}\tilde{v}_2;\ \ \tilde{v}_4=-
\frac{\bar{D}_3}{\bar{D}_0}\tilde{r}_e-
\frac{\bar{D}_2}{\bar{D}_0}\tilde{v}_2-
\frac{\bar{D}_1}{\bar{D}_0}\tilde{v}_3;\\
&& \tilde{a}^{-1}\equiv \frac{M}{4\pi a}-\bar{J}_0,\ \
\tilde{r}_e\equiv \frac{M}{8\pi} r_e,\ \  \tilde{v}_n\equiv
\frac{M}{4\pi}v_n, n\geq2,\eea where the coefficients
$[\frac{\bar{D}_k}{\bar{D}_0} ]$ could be solved in terms of
$M,a,r_e,v_2,v_3$ and $v_4$.
\subsection{Lessons from nonperturbative solutions}Now it is clear
that things get more complicated as more higher order terms are
included in the potential. Given this, we should not take the
lower order results too seriously. For instance, the low energy
theorems at leading order are too simple to be true in practice:
$r_e=v_k=0, k\geq2$. This implies the necessity to include higher
order terms, which, however, will bring us both favorable and
unfavorable consequences. On one hand, more severe prescription
dependence will show up and make this analysis more difficult. On
the other hand, more prescription ambiguities also provide us with
more chances to access the measured values of the ERE parameters.
Although our calculations were done for the case of contact
potentials, the core feature of our analysis--more ambiguities or
more divergences at higher orders--holds true also for realistic
potentials. At this stage, we shall mention that the freedoms in
the prescription are in fact limited: $[\bar{J}_n]$ must satisfy
certain requirements as presented in the discussion following
Eq.(\ref{J_para}). Moreover, the coupling constants should
generally follow certain rules of EFT power counting. Then, after
putting all these theoretical aspects into consideration, the EFT
predictions must lie in certain region of the 'space' of
observables.

Now we provide another way to see the virtue of the fitting
procedure.  Let us examine the variation of the functional form of
the scattering length $a$ in terms of the couplings and
$[\bar{J}_n]$ for different chiral orders:\bea &&\Delta=0:\ \
a=\frac{M}{4\pi}\frac{C_0}{1+C_0\bar{J}_0},\nonumber \\
&&\Delta=2:\ \ a=\frac{M}{4\pi}\frac{C_0+C_2^2\bar{J}_5}{(1-C_2
\bar{J}_3)^2+\bar{J}_0(C_0+C_2^2\bar{J}_5)},\nonumber \\
&&\Delta=4:\ \ a=\frac{M}{4\pi}\frac{\bar{D}_0
(C_0,C_2,C_4,\tilde{C}_4;\bar{J}_5,\bar{J}_7,\bar{J}_9)}
{\bar{N}_0(C_0,C_2,C_4,\tilde{C}_4;\bar{J}_3,\bar{J}_5,\bar{J}_7,\bar{J}_9)+
\bar
{J}_0\bar{D}_0(C_0,C_2,C_4,\tilde{C}_4;\bar{J}_5,\bar{J}_7,\bar{J}_9)}.
\nonumber\eea  It is obvious that the theoretical form of the
scattering length varies with the chiral order quite
significantly! So the scattering length calculated at lower orders
should not be directly identified with the experimental value in
order to accommodate the higher order terms. Thus a more
reasonable way for fixing the renormalization prescription will be
the one that avoids the direct identification of physical
parameters in order to accommodate higher order contributions in a
consistent way. To this end, again a procedure like fitting the
empirical curve over appropriate low energy regions might be more
plausible.

\section{power counting and renormalization in the nonperturbative regime}
In all the discussions above, we have left out the power counting
of the couplings. Since they constitute the basis for the EFT
methods, it is necessary to see what the nonperturbative
renormalization procedure described above means for the power
counting rules. In fact, as was stressed in Ref.\cite{JFY}, the
parametrization in the nonperturbative regime given in
Eq.(\ref{para}) implies that, in order for a power counting scheme
for couplings to be meaningful, the corresponding prescription for
the constants $[\bar{J}_n]$ must be appropriately chosen.
Otherwise, one could not obtain the physical $T$-matrix.

From the standpoint of UT, both $[C_n]$ and $[\bar{J}_n]$ come
from the well defined low energy projection
($\breve{{\mathcal{P}}}_{\text{\tiny LE}}$) applied to UT
amplitudes. So both $[C_n]$ and $[\bar{J}_n]$ serve as the
elementary parameters for parametrizing the $T$-matrix for the low
energy nucleon-nucleon scattering. In the EFT treatment without
knowledge of the details from UT, we are forced to employ $[C_n]$
as the elementary couplings according to certain counting rules,
while the constants $[\bar{J}_n]$ appear as the divergent pieces
in the EFT loops constructed with the use of $[C_n]$. Thus it is
the EFT treatment that makes $[C_n]$ and $[\bar{J}_n]$ look
disparate. In UT they are organized and derived together according
to more elementary rules. Therefore, changing any of them (each
single parameter in $[C_n]\bigcup[\bar{J}_n]$) alone would alter
the physical behavior of $T$. Thus they must be considered
together.

One could also understand it from the Wilsonian definition of EFT
through successive decimation of the higher scales, where
different EFT expansion point would lead to both different
couplings and different $[J_{n}]$, as long as the expansions are
compatible with the chiral power counting.

To be specific, the variations of $[C_n]$ and $[\bar{J}_n]$ (from
now on $\bar{J}_0$ is excluded from $[\bar{J}_{\ldots}]$ for the
reasons to be given below) must not alter the functional form
(shape) of the $T$-matrix:\bea \texttt{\large Re}\left
\{\frac{1}{T_{os}(p)}-\bar{J}_0\right\}= \frac{\sum
\bar{N}_i([C_{\ldots}^{\prime}];[\bar{J}_{\ldots}^{\prime}])
p^{2i}}{\sum
\bar{D}_j([C_{\ldots}^{\prime}];[\bar{J}_{\ldots}^{\prime}])p^{2j}}=
\frac{\sum \bar{N}_i([C_{\ldots}];[\bar{J}_{\ldots}]) p^{2i}}{\sum
\bar{D}_j([C_{\ldots}];[\bar{J}_{\ldots}])p^{2j}}= \frac{\sum
N_i^{(\texttt{\tiny phys})} p^{2i}}{\sum D_j^{(\texttt{\tiny
phys})}p^{2j}}.\eea Here we use the superscript '$\texttt{phys}$'
to indicate that the parameters in the last fraction are
physically determined, for example, from a genuine UT. To see why
$\bar{J}_0$ is excluded from $[\bar{J}_n]$, consider the physical
parametrization of $T$-matrix (independent of the variations of
$[C_{\ldots}]$ and $[\bar{J}_n]$), which has the following form
\bea \frac{1}{T^{(\texttt{\tiny phys})}}=\frac{\sum
N_i^{(\texttt{\tiny phys})} p^{2i}}{\sum D_j^{(\texttt{\tiny
phys})}p^{2j}}+ M\gamma+\frac{M}{4\pi}ip, \eea where $\gamma$ must
be a physical scale, just like the nucleon mass $M$ and the
on-shell momentum $p$. Now it is clear that $\bar{J}_0$ alone
corresponds to the physical parameter $M\gamma$, which should
therefore be independent of prescriptions. If it were not the
case, or if $\bar{J}_0$ could vary with prescriptions, we would
have to alter $\bar{N}_0$ and $\bar{D}_0$ to compensate for such
variation in $\bar{J}_0$. Now, to keep the proportionality between
$\bar{N}_0,\bar{D}_0$ and $\bar{N}_{i}p^{2i},\bar{D}_{j}p^{2j}$
invariant, all the rest of $[\bar{N}_{\ldots},\bar{D}_{\ldots}]$
must be accordingly altered, which in turn leads to an overall
factor for $(\sum N_i^{(\texttt{\tiny phys})} p^{2i})/(\sum
D_j^{(\texttt{\tiny phys})}p^{2j})$. Then, the functional
dependence of the $T$-matrix upon $p$ would be altered, since its
imaginary part, $\frac{M}{4\pi}ip$, remains intact. Hence
$\bar{J}_0$ must stay independent of prescriptions, i.e.,
physical. One could also verify this by examining the consequences
on the ERE parameters.

As $p$ is arbitrary in the supposed range, the 'invariance'
discussed above leads to the following nontrivial equations for
$[C_n]$ and $[\bar{J}_n]$ with the crucial presence of the
physical parameters $[ N_i^{(\texttt{\tiny phys})}]$ and
$[D_j^{(\texttt{\tiny phys})}]$ for the on-shell $T$-matrix, \bea
\label{preRGE}
\bar{N}_i([C_{\ldots}];[\bar{J}_{\ldots}])=N_i^{(\texttt{\tiny
phys})},\ \ \
\bar{D}_j([C_{\ldots}];[\bar{J}_{\ldots}])=D_j^{(\texttt{\tiny
phys})},\ \ \forall i,j.\eea These equations have dual
implications: they could be used either (1) to fix the
prescription ($[\bar{J}_{\ldots}]$) in terms of the couplings
($[C_{\ldots}]$) and the physical parameters
($[N_{\ldots}^{(\texttt{\tiny phys})};D_{\ldots}^{(\texttt{\tiny
phys})}]$) or conversely (2) to examine the influence of
prescription upon the couplings with the help of the physical
parameters. The first use just parallels what we have done in Sec.
IV.B and C.

\subsection{Interplay between Power counting and
prescription: next-to-leading order} Let us illustrate the
interplay between power counting and prescription at
next-to-leading order; that is, we try to solve the following
equations for couplings:\bea\label{preRGE2}
\frac{C_0+C^2_2\bar{J}_5}{(1-C_2\bar{J}_3)^2}=\alpha_0
\equiv\frac{D^{(\texttt{\tiny phys})}_0}{N^{(\texttt{\tiny
phys})}_0},\ \ \
\frac{2C_2-C^2_2\bar{J}_3}{(1-C_2\bar{J}_3)^2}=\alpha_2
\equiv\frac{D^{(\texttt{\tiny phys})}_2}{N^{(\texttt{\tiny
phys})}_0}.\eea The solutions are easy to find as,\bea
\label{C_2solution}&&C_2^{(\pm)}=\bar{J}_3^{-1} \{ 1 \pm
(1+\alpha_2\bar{J}_3)^{-\frac{1}{2}} \},\\
\label{C_0solution}&&C_0^{(\pm)}=\frac{\alpha_0}{1+\alpha_2\bar{J}_3}-
\frac{\bar{J}_5}{\bar{J}_3^2} \{ 1 \pm
(1+\alpha_2\bar{J}_3)^{-\frac{1}{2}} \}^2.\eea Taking into account
the natural boundary condition for $C_2$: $C_2|_{J\rightarrow
0}\Longrightarrow \alpha_2/2$, we are left with unique solution:
$C_2^{(-)}$ (and $C_0^{(-)}$). Thus assigning a power counting to
$C_0$ and $C_2$ means assigning the sophisticated scaling for
$\bar{J}_3$ and $\bar{J}_5$. Conversely, one can come up with an
alternative interpretation: The power counting for the couplings
could only be preserved in some particular prescription in order
to obtain the expected physical behavior from the $T$-matrix. Note
that here we have deliberately not mentioned $\bar{J}_0$, it will
be exclusively discussed below. Equations (\ref{preRGE}) or
(\ref{preRGE2}) now formalize our discussions concerning the
interplay between power counting and prescription.

More interestingly, these equations have a further utility: they
could be used to describe the evolution of the couplings in terms
of a sliding scale ($\mu$) in $[\bar{J}_n](=[q_n M \mu^n,
n\neq0])$. Since the exogenous counter terms are incompatible with
the closed form of the $T$-matrix, the conventional route to the
evolution described by renormalization group equation does not
exist. But we could take the evolution implied by Eqs.
(\ref{preRGE}) or (\ref{preRGE2}) as a nonperturbative
"renormalization group" evolution. We discuss this point in the
next subsection.

\subsection{Nonperturbative 'renormalization group' (RG) evolution}
To proceed, let us choose the prescription with $[\bar{J}_n\equiv
q_nM\mu^n]$ to examine the evolution of the couplings enforced by
(\ref{preRGE2}). Let us assume that there exist enough boundary
conditions to obtain the 'physical' solutions for the couplings
from the equations in (\ref{preRGE}):\bea\label{RGE}
C_i=F_i([N_{\ldots}^{(\texttt{\tiny
phys})},D_{\ldots}^{(\texttt{\tiny phys})},M];[q_{\dots}];\mu),
\forall i.\eea With such nonperturbative solutions, the complete
evolution of the couplings are determined and both the IR and the
UV fixed points can be identified. For example, at next-to-leading
order, we have from Eqs.(\ref{C_0solution},\ref{C_2solution}),\bea
&&C_0(\alpha_0,\alpha_2,M,q_3,q_5;\mu)=\frac{\alpha_0}{1+q_3\alpha_2M\mu^3}-
\frac{q_5}{q_3^2M\mu} \{ 1 -
(1+q_3\alpha_2M\mu^3)^{-\frac{1}{2}} \}^2,\\
&&C_2(\alpha_0,\alpha_2,M,q_3,q_5;\mu)=(q_3M\mu^3)^{-1} \{ 1 -
(1+q_3\alpha_2M\mu^3)^{-\frac{1}{2}}\}. \eea It is easy to see
that they have both IR and UV fixed points:
\bea\label{fixedpoint}&&\texttt{IR fixed point}(\mu\Rightarrow0):
\ \ C_0^{(IR)}=\alpha_0,\ \ C_2^{(IR)}=\alpha_2/2;\\
&&\texttt{UV fixed point}(\mu\Rightarrow\infty):\ \
C_0^{(UV)}=C_2^{(UV)}=0.\eea Note that the prescription dependence
is obvious in Eq.(\ref{RGE}) with the presence of $[q_{\ldots}]$,
but the UV and IR fixed points are prescription independent. While
the IR fixed points are realistic as the couplings were defined in
the low energy limit, the UV fixed points seem not to be
realistic. But such UV behavior of the EFT couplings is compatible
with the fact that the EFT couplings would be dominated by the UT
couplings at high energy, and therefore 'vanish'. Of course we
should bear in mind that, what we obtained are only approximate
answers, though nonperturbative.

Note that the Eqs.(\ref{preRGE},\ref{RGE}) contain the full
dependence upon the prescription parameters. So one could also
derive the equations a la St\"uckelberg and Petermann\cite{SP}
that describe the laws for transitions from one prescription to
another which are not related by running the renormalization
scale:\bea \label{shapeRG} \frac{d}{d[\bar{J}_{\ldots}]}\left
\{\bar{N}_i,\bar{D}_j\right \}=0, \forall i,j. \eea In terms of
$[q_{\ldots};{\mu}]$ they become \bea \frac{d}{d[q_{\ldots}]}\left
\{\bar{N}_i,\bar{D}_j\right \} =0, \forall i,j. \eea

In the foregoing discussions the physical requirements are imposed
on the functional shape of the $T$-matrix. Alternatively, we could
also employ the physically determined ERE parameters (scattering
length, effective range,etc) instead of
$[N_{\ldots}^{(\texttt{\tiny phys})};D_{\ldots}^{(\texttt{\tiny
phys})}]$ to solve the couplings in terms of $[\bar{J}_n]$. In
principle, the two approaches should lead to the same evolution
behavior, but the ERE approach is more involved than the shape
approach as is clear from the comparison between the Eqs.
(\ref{ERE2},\ref{ERE4}) and Eqs.(\ref{preRGE}).

\subsection{Determinant for the natural or unnatural scattering length:
$\bar{J}_0$} Now we discuss the determinant(s) of the size of
physical parameters $[ N_i^{(\texttt{\tiny phys})}]$ and
$[D_j^{(\texttt{\tiny phys})}]$ or $a, r_e, v_k, \forall k\geq2$.
As argued above, a complete parametrization of the $T$-matrix is
given by $[C_{\ldots}]$ supplemented with $[\bar{J}_{\ldots}]$,
then $[C_{\ldots}]$ and $[\bar{J}_{\ldots}]$ together determine
whether the physical parameters are of natural size or not. We
could have four rough scenarios, listed in Table \ref{table},
where by a natural $C_{n}$ we mean that the scale $\Lambda$ in its
parametrization $C_n\sim 1/(M\Lambda^{n+1})$ is of the size of the
expansion scale (unnatural if $\Lambda\sim p,m_{\pi}$), while for
$[\bar{J}_{\ldots}=q_{\ldots}M\mu^{\ldots}]$ the situation is
reversed: the natural size of $\mu$ should be $\sim p, m_{\pi}$. A
natural $T$-matrix is parametrized by $[ N_i^{(\texttt{\tiny
phys})},D_j^{(\texttt{\tiny phys})}]$ (or for $a, r_e, v_k,
\forall k\geq2$) such that the dimensional parameters are of the
same magnitudes as the natural couplings.

Examining the concrete expressions of the $T$-matrix, we find that
whether the $T$-matrix is natural or not is determined by both the
sizes of the couplings and the magnitudes of the dimensionless
combinations like $\prod_{n,m} C_n^{\pm 1}\bar{J}_{m}$
($\texttt{dim}[\prod_{n,m} C_n^{\pm 1}\bar{J}_{m}]=0$). Now
suppose we have natural couplings, i.e., $C_{n}\sim
1/(M\Lambda^{n+1}), \forall n$. If $[\bar{J}_{\ldots}]$ are also
natural, we should have $|\prod_{n,m} C_n^{\pm 1}\bar{J}_{m}|\ll1$
for all the dimensionless combinations. Then, given our experience
at next-to-leading and next-to-next-to-leading orders, we can
anticipate that: \bea \label{natural} N^{\texttt{\tiny
(phys)}}_0\sim1,\ \ N^{\texttt{\tiny (phys)}}_i \sim
\frac{1}{\Lambda^{2i}};\ \ D^{\texttt{\tiny (phys)}}_j\sim
\frac{1}{M\Lambda^{j+1}}(\sim C_j), \forall i,j.\eea In such case
we shall obtain a natural $T$-matrix, or ERE parameters
($a,r_e,v_k,\forall k\geq2$), of natural sizes. If
$[\bar{J}_{\ldots}]$ are unnatural, then in general, we could have
$|\prod_{n,m} C_n^{\pm 1}\bar{J}_{m}|\sim1$ for the dimensionless
combinations. Therefore, we
have,\bea\label{unnatrual}N^{\texttt{\tiny (phys)}}_0 \gg1 (\ll1)
,\ \ N^{\texttt{\tiny (phys)}}_i \gg
 (\ll)\frac{1}{\Lambda^{2i}};\ \ D^{\texttt{\tiny (phys)}}_j\gg
 (\ll)
\frac{1}{M\Lambda^{j+1}}(\sim C_j), \forall i,j. \eea In this
case, we obtain an unnatural $T$-matrix, or  unnatural ERE
parameters ($a,r_e,v_k,\forall k\geq2$) with natural couplings.
For example, at next-to-leading order, we have,\bea
\texttt{natural}\ [\bar{J}_{\ldots}]:&&
|C_2\bar{J}_3|\ll1, |C_2^2C_0^{-1}\bar{J}_5|\ll1,\nonumber \\
\Longrightarrow \texttt{natural
$T$}:&&\frac{(1-C_2\bar{J}_3)^2}{C_0+C_2^2
\bar{J}_5+C_2(2-C_2\bar{J}_3)p^2} \simeq \frac{1}{C_0+2C_2p^2};\\
\texttt{unnatural}\ [\bar{J}_{\ldots}]: &&|C_2\bar{J}_3|\sim1,
|C_2^2C_0^{-1}\bar{J}_5|\sim1,\nonumber\\
\Longrightarrow \texttt{unnatural
$T$}:&&\frac{(1-C_2\bar{J}_3)^2}{C_0+C_2^2
\bar{J}_5+C_2(2-C_2\bar{J}_3)p^2} =
\frac{\zeta_1}{\zeta_2C_0+2\zeta_3C_2p^2},\eea where each of
$\zeta_{\ldots}$ can be either pretty small or pretty large and
therefore the $T$-matrix could not be a natural one.

Now let us consider $\bar{J}_0$ and the scattering length in
particular. As argued above, $\bar{J}_0$ should be viewed as an
independent physical parameter, not as a common prescription
parameter. From the parametrization of $T$ and the formulae in the
preceding sections, $\bar{J}_0$ will only contribute to the
scattering length:\bea &&a^{-1}=\frac{4\pi}{M}\left
\{\bar{J}_0+\frac{\bar{N}_0([C_{\ldots},\bar{J}_{\ldots}])}
{\bar{D}_0([C_{\ldots},\bar{J}_{\ldots}])}\right \},\nonumber \\
&&r_e=\frac{8\pi}{M}\left \{
\frac{\bar{N}_0([C_{\ldots},\bar{J}_{\ldots}]) \bar{D}_1
([C_{\ldots},\bar{J}_{\ldots}])}{\bar{D}_0^2([C_{\ldots},\bar{J}_{\ldots}])}
 -\frac{\bar{N}_1 ([C_{\ldots},\bar{J}_{\ldots}])}{
 \bar{D}_0 ([C_{\ldots},\bar{J}_{\ldots}])}
 \right\}, \cdots.\nonumber \\
\Rightarrow&&\frac{\partial a^{-1}}{\partial
\bar{J}_0}=\frac{4\pi}{M},\ \ \frac{\partial r_e}{\partial
\bar{J}_0}=\frac{\partial v_k}{\partial \bar{J}_0}=0, \forall
k\geq2.\eea Now we can see that, even when both $[C_{\ldots}]$ and
$[\bar{J}_{\ldots}]$ are of natural sizes, the scattering length
could be unnaturally large once $\bar{J}_0$ is unnatural ($\sim
M\Lambda$),\bea &&\texttt{natural} \bar{J}_0 (\sim M\mu):
 a^{-1} \simeq
-{\mathcal{O}}(\Lambda) +{\mathcal{O}}(\mu)\sim
-{\mathcal{O}}(\Lambda);\\
&&\texttt{unnatural} \bar{J}_0(\sim M\Lambda): a^{-1} \simeq
-{\mathcal{O}}(\Lambda) +{\mathcal{O}}(\Lambda)\sim
-{\mathcal{O}}(\mu).
 \eea That is, in the $^1S_0$ channel, there
theoretically exists such a scenario that the scattering length
could be unnaturally large while all the rest ERE parameters are
naturally sized. Then the first situation in Table \ref{table}
should be amended as follows: even when all the couplings and all
the rest $[\bar{J}_{\ldots}]$ are natural, we would get an
unnatural scattering length as long as $\bar{J}_0$ is unnatural.

For an unnatural power counting of the couplings, the discussion
would be more difficult and we refrain from exploring such
situations here. As we have shown, that both the natural and the
unnatural physical parameters could be explained with the natural
couplings (provided the nontrivial nonperturbative prescription
dependence is fully explored), we feel that it is more reasonable
to work with natural or conventional power counting of EFT
couplings.

\section{Discussions and Summary}

Now it is time for us to address some theoretical aspects that
have been omitted or not fully discussed so far. Let us start with
the relation between the UT renormalization and the EFT
renormalization that is involved in Sec. III.C. Generally, in an
EFT one deals with the new divergences in the diagrams that are
induced by some low energy expansion (with the "wrong" order of
operations, C.f. Sec. III.C.) or similar operation in UT, whereas
the diagrams that need renormalization in UT are usually hidden in
the EFT couplings. Note, that the diagrams that renormalize UT
dominate the quantum fluctuations at short distances, while the
ones that are divergent in EFT dominate those at long distances.
So the two renormalization do not interfere with (affect) each
other due to the large scale hierarchy between UT and EFT, i.e.,
they work at two widely separated scales. Thus, the
renormalization in UT does not affect the renormalization in EFT.
This supplements our remarks after Eq.(\ref{Tn3r}) in Sec. III.C.

Next, let us address the effect of the potential truncation on the
nonperturbative renormalization group evolution. At any fixed
chiral order, the nonperturbative evolution behaviors of the
coupling of the highest chiral dimension should be less
trustworthy. This is because once the next order interactions are
included, the coefficients for the term with the highest power of
$p$ would suffer the largest changes in the functional forms; the
coefficients for the lower power terms receive smaller changes
from the new couplings. That means that due to the truncation of
the potential, the nonperturbative evolution behaviors of the EFT
couplings with lower chiral dimensions should be more trustworthy
than those with higher chiral dimensions. One could see this point
by noting how the forms of $[N_i([C_{\ldots}];[\bar{J}_{\ldots}]),
D_i([C_{\ldots}];[\bar{J}_{\ldots}])]$ (as functions of the
couplings $[C_{\ldots}]$) change with the inclusion of higher
order interactions.

In Sec. V.C., we have shown that a natural (or conventional)
chiral power counting of the EFT couplings does allow the
$T$-matrix to have unnatural parameters, or unnatural scattering
length, etc. In particular, there is a possibility that only the
scattering length is unnatural while the rest of the parameters
are natural. This seems to be just the realistic situation with
the $^1S_0$ channel nucleon-nucleon scattering at low energy. This
scenario is clearly different from the one discussed in the
literature where unusual power counting of the couplings was
employed\cite{KSW}. Here the key role is played by the
nonperturbative renormalization prescription.

Although our conclusions or remarks have been reached with contact
interactions, we feel that the conclusions or scenarios depicted
here should remain qualitatively true even in a realistic
situation because the crucial features of the nonperturbative
renormalization remain unchanged: (1) More ill defined pieces in
the loop integrals appear at higher chiral orders; (2) The
nonperturbative solution of the $T$-matrix takes a closed form
that can only be renormalized via endogenous counter terms.
Alternatively, one could also take the rational function form as a
Pad\'e approximant to the realistic $T$-matrix.

Now let us comment on the literature. In Ref.\cite{BBSvK}, a
subtraction similar to the endogenous one described in the present
paper is employed: the counter term is introduced before the
$T$-matrix is calculated, a procedure that parallels the loop
integrations. However, it is not clear if the subtraction
described in some papers is equivalent to the endogenous one or
not. For example, the subtraction procedure described in
Ref.\cite{Frederico} does not appear to be an endogenous one. Thus
it may be flawed, as was already noted in Ref.\cite{Soto}. In
Ref.\cite{Soto}, the whole investigation is made in the
nonperturbative formulation (compact) of the $T$-matrix, a
positive aspect of this study. However, a special regularization
(cutoff regularization) exclusively used in Ref.\cite{Soto}
unfortunately makes their analysis inevitably prescription
dependent. In a contrast, the strategy employed in
Ref.\cite{Nieves} for parametrizing and fixing the nonperturbative
renormalization prescription dependence is closer to the one used
in the present paper. The importance of boundary conditions has
already been stressed in Ref.\cite{VA}, where the physical
observables, such as phase shifts, were parametrized without
involving explicit divergences.

Obviously, we just explored some convenient scenarios of the
nonperturbative solutions. Our arguments have been unable to
exclude many other possible scenarios. The only point in favor of
the scenarios discussed in this paper is that they are relatively
simple, whereas the rest possibilities seem rather sophisticated,
and often use fine tuning or similar arguments.

In our opinion, a better way to work with the renormalization in
the nonperturbative regime is to appreciate the presence of a
well-defined theory underlying an EFT, as illustrated in this
paper. In this sense, the renormalization of singular potentials
in quantum mechanics\cite{Jackiw}, or equivalently the
self-adjoint extension of singular operators in Hilbert space,
should also be embedded in the underlying theory background. This
is plausible since quantum mechanics IS an effective theory of
quantum field theory.

In summary, we reconsidered the renormalization of the EFT for
nucleon-nucleon scattering in the nonperturbative regime using
contact potentials that facilitate rigorous solutions of LSE.
Detailed analysis reveals that the $T$-matrix in the
nonperturbative regime should be renormalized through the
endogenous counter terms whose net effects are to remove the
divergences in the loop integrals, or through means that could
yield the same results. The rationality for the subtractions at
loop integral level could be naturally explained from the
underlying theory, with the UV divergences being shown to come
from the "incorrect" order of operations in the construction of
EFT. Then, using the effective range expansion, we demonstrated
that the nontrivial renormalization prescription dependence in the
nonperturbative regime must be "removed" by imposing appropriate
boundary conditions. We also argued that when imposing boundary
conditions, the full "space" for renormalization prescriptions
should be explored in order to be able to remove any residual
prescription dependence. It is also important to impose the
boundary conditions in such a way that higher order terms in the
potential could be consistently incorporated. Finally, the
nontrivial relation between the power counting of the couplings
and the renormalization prescription was highlighted in the
nonperturbative regime. As byproducts, (1) the nonperturbative
'renormalization group' evolution was described; (2) the
naturalness of the scattering length, etc. were shown to be
compatible with the natural or conventional power counting of the
couplings because of the nontrivial prescription dependence. That
is, the nontrivial prescription dependence becomes a virtue in
such a case. Obviously, much work remains to be done.

\begin{table}[t]
\caption{Determinant for naturalness/unnaturalness of $T$-matrix}
\begin{center}
\begin{tabular}{c|c|c} \hline
 & Natural $[C_{\ldots}]$ & Unnatural $[C_{\ldots}]$ \\ \hline
Natural $[\bar{J}_{\ldots}]$& Natural $T$-matrix &Unnatural $T$-matrix \\
\hline Unnatural $[\bar{J}_{\ldots}]$& Unnatural $T$-matrix &
Natural/Unnatural $T$ ?\\
\hline
\end{tabular}
\end{center}
\label{table}
\end{table}
\section*{Acknowledgement}
It is a pleasure for one of the authors (J.-F. Yang) to
acknowledge the valuable discussions with Professor W. Zhu (ECNU,
Shanghai), Professor Xiao-yuan Li (ITP, CAS, Beijing) and
Professor Guang-lie Li (IHEP, CAS, Beijing) on EFT and nuclear
forces. He is also grateful to the editors for their very kind
helps. Special thanks are due to Dr. A. Onishchenko (Uni.
Hamburg) for his critical reading of the manuscript that greatly
improved the presentation. The project is supported by the
National Natural Science Foundation under Grant No. 10502004.

\section*{Appendix A}
\begin{eqnarray}
\label{I}{\mathcal{I}}(E^+)\equiv\left(%
\begin{array}{ccc}
  \int \frac{d^3k}{(2\pi)^3}\frac{1}{E^+-\frac{k^2}{M}}, &
  \int \frac{d^3k}{(2\pi)^3} \frac{k^2}{E^+-\frac{k^2}{M}},&
  \int \frac{d^3k}{(2\pi)^3} \frac{k^4}{E^+-\frac{k^2}{M}}
   \\
  \int \frac{d^3k}{(2\pi)^3} \frac{k^2}{E^+-\frac{k^2}{M}}, &
  \int \frac{d^3k}{(2\pi)^3} \frac{k^4}{E^+-\frac{k^2}{M}},&
  \int \frac{d^3k}{(2\pi)^3} \frac{k^6}{E^+-\frac{k^2}{M}} \\
\int \frac{d^3k}{(2\pi)^3} \frac{k^4}{E^+-\frac{k^2}{M}},  & \int
\frac{d^3k}{(2\pi)^3} \frac{k^6}{E^+-\frac{k^2}{M}},&
\int \frac{d^3k}{(2\pi)^3} \frac{k^8}{E^+-\frac{k^2}{M}}\\
\end{array}%
\right).
\end{eqnarray} The entries of this
matrix can be parametrized as follows ($p\equiv\sqrt{M E}$):
\begin{eqnarray}
\label{integrals}&& \int
\frac{d^3k}{(2\pi)^3}\frac{1}{E^+-\frac{k^2}{M}}\equiv
I_0=-J_0-i\frac{M p}{4\pi};\\&& \int \frac{d^3k}{(2\pi)^3}
\frac{k^2}{E^+-\frac{k^2}{M}}\equiv J_3+ I_0p^2;\\ &&\int
\frac{d^3k}{(2\pi)^3} \frac{k^4}{E^+-\frac{k^2}{M}}\equiv
J_5+J_3p^2+I_0p^4;\\&&\int \frac{d^3k}{(2\pi)^3}
\frac{k^6}{E^+-\frac{k^2}{M}}\equiv
J_7+J_5p^2+J_3p^4+I_0p^6;\\&&\int \frac{d^3k}{(2\pi)^3}
\frac{k^8}{E^+-\frac{k^2}{M}}\equiv
J_9+J_7p^2+J_5p^4+J_3p^6+I_0p^8.
\end{eqnarray}Here $\{J_n\}$ with $n=0,3,5,7,9$ are regularization
and renormalization prescription dependent constants.
\section*{Appendix B}
\begin{eqnarray}
N_0=&&(1-C_2J_3-C_4J_5)^2-C_0\tilde{C}_4J_3^2- \tilde{C}_4
J_5+2\tilde{C}_4 C_4J_5^2-\tilde{C}_4
C_4^2J_5^3-2\tilde{C}_4 C_4 J_3J_7\nonumber \\
&&-\tilde{C}_4
C_4^2J_3^2J_9+2\tilde{C}_4C_4^2J_3J_5J_7;\nonumber\\
N_1=&&-2C_4 J_3-\tilde{C}_4 J_3 +2 C_2C_4 J_3^2+2\tilde{C}_4 C_4
J_3 J_5+2C_4^2 J_3J_5-\tilde{C}_4C_4^2 J_3J_5^2+\tilde{C}_4C_4^2
J_3^2 J_7 ;\nonumber \\N_2=&& C_4^2J_3^2;\nonumber \\
D_0=&&C_0+C_2^2J_5+C_4^2J_9-C_0\tilde{C}_4 J_5
+C_4^2\tilde{C}_4 J_7^2+2C_2C_4J_7-C_4^2\tilde{C}_4J_5J_9 ;\nonumber \\
D_1=&&2C_2-C_2^2J_3+C_0\tilde{C}_4J_3+C_4^2J_7 +2C_4\tilde{C}_4
J_7-C_4^2\tilde{C}_4 J_5 J_7+\tilde{C}_4C_4^2J_3J_9; \nonumber\\
D_2=&&2C_4+\tilde{C}_4-2C_2C_4J_3-2C_4\tilde{C}_4 J_5-C_4^2J_5+
\tilde{C}_4 C_4^2J_5^2-C_4^2\tilde{C}_4 J_3J_7; \nonumber
\\D_3=&&-C_4^2J_3.\\
J^{\text{\tiny (UT)}}_0(M,m_h)=&&\frac{m_h^4}{(4\pi)^2M^2}\int^1_0
dx \int^x_0 dy
\frac{(y+3-2x)^2+8(x-1)}{[(y+1-2x)^2+y\frac{m_h^2}{M^2}]^2}.
\end{eqnarray}
\section*{Appendix C}
Consider the contact potential given at any chiral order. In the
matrix form defined in Sec. III.A, we have $V=U^T\lambda
U,T=U^T\tau U$, with $U(p)\equiv (1,p^2,p^4,p^6,\ldots)$ being a
column vector and $U^T$ being the transposed vector. Then the
convolution in LSE could be factorized as $ VG_0T=U^T(p)\lambda
{\mathcal{I}}\tau U(q)$, with the matrix ${\mathcal{I}}$ being
defined as follows, \bea {\mathcal{I}}\equiv \int
\frac{kdk^2}{(2\pi)^2}\frac{U(k)U^T(k)} {E-k^2/M+i\epsilon}.\eea
The 3$\times$3 case of ${\mathcal{I}}$ is given in Appendix A. It
is easy to see that we could rewrite ${\mathcal{I}}$ as follows,
\bea\label{complex_I} {\mathcal{I}}=I_0 U(\sqrt{ME})U^T(\sqrt{ME})
+\tilde{{\mathcal{I}}}([J_m];ME), m\neq0,\eea where $I_0$ and
$J_m$ with $m\neq0$ are defined in Appendix A. Here
$\tilde{{\mathcal{I}}}$ is a real matrix independent of $I_0$.
From Eq.(\ref{complex_I}) it follows that, \bea VG_0T=U^T\lambda
(I_0UU^T+\tilde{{\mathcal{I}}})\tau U =I_0VT+
U^T\lambda\tilde{{\mathcal{I}}}\tau U\eea Then, using the
parametrization in (\ref{npt}), we find that, for on-shell
momentum: \bea
\label{para_C}T^{-1}=V^{-1}-{{\mathcal{G}}}=V^{-1}-I_0-
\tilde{{\mathcal{G}}},\eea with $\tilde{{\mathcal{G}}}\equiv
\frac{U^T\tilde{\lambda} \tilde{{\mathcal{I}}} \tau
U}{\tilde{V}T}={\mathcal{G}}-I_0$. Now comparing this with the
following representation of $T$ derived in Ref.\cite{JFY} using
the relation between the on-shell $T$-matrix and on-shell
$K$-matrix, $ T^{-1}=K^{-1}+ \frac{M}{4\pi}ip$, we could find
that, \bea\tilde{{\mathcal{G}}}=V^{-1}-K^{-1}+J_0,\eea that is,
$\tilde{{\mathcal{G}}}$ must be a real number. But this real
quantity is constructed with a complex $T$ that contains the
infinite iterations of the complex number $I_0$ as given in
(\ref{para_C}). That means $I_0$ must cancel out in the infinite
iteration, and hence must disappear in the real quantity
$\tilde{{\mathcal{G}}}$. This in turn implies that, $J_0$, as the
real part of $I_0$, does not appear in $\tilde{{\mathcal{G}}}$.
Finally these facts will lead to following from of $T$-matrix
constructed with local potential:\bea T^{-1}=\frac{\sum
N_i([C_{\ldots}],[J_{\ldots}])p^{2i}} {\sum
D_j([C_{\ldots}],[J_{\ldots}])p^{2j}}-I_0=\frac{\sum
N_i([C_{\ldots}],[J_{\ldots}])p^{2i}} {\sum
D_j([C_{\ldots}],[J_{\ldots}])p^{2j}}+J_0+ \frac{M}{4\pi}ip,\eea
with $[N_i,D_j]$ being independent of $J_0$. $QED$.

\begin{figure}[h]
\begin{center}
\includegraphics*[width=20cm]{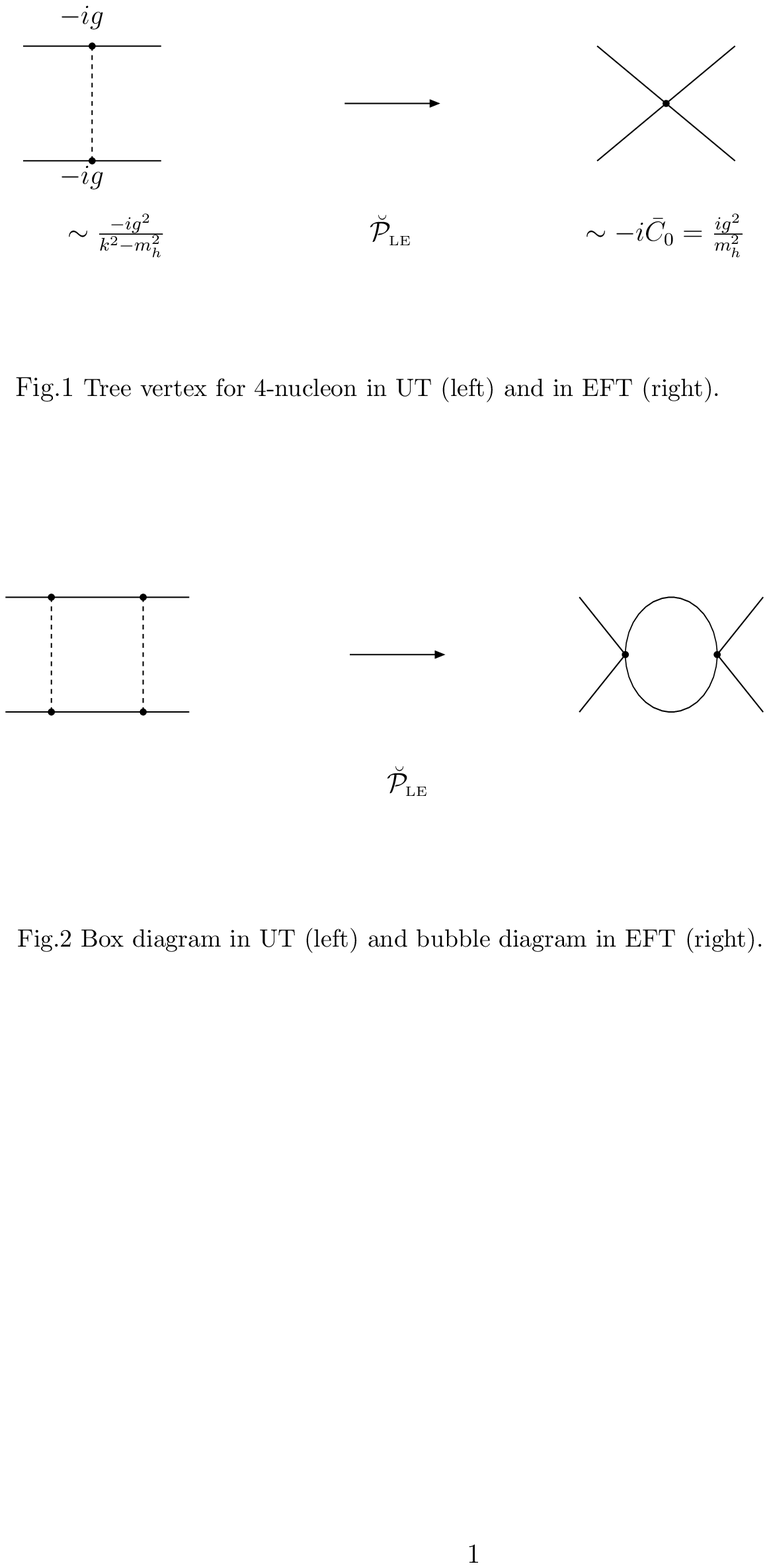}
\end{center}
\end{figure}

\end{document}